A photon observed by the SEDA-FIB detector on the arrival of the gravitational wave

Y. Muraki (1),   K. Kamiya(2),   K. Koga(2),   H. Matsumoto(2), and S. Shibata(3)

 1) Inst for Space Earth Environment, Nagoya University, Chikusa, Nagoya, Japan
 2) Space environment monitor team, Tsukuba space center, JAXA, Tsukuba, Japan
 3) Faculty of Engineering, Chubu University, Kasugai, Aichi, Japan

Abstract

We reported a photon detection at the arrival time of the gravitational wave on December 26th, 2015.   According to the LIGO-Virgo collaboration, the gravitational wave was produced by the merging process of the two black holes.   The merged time was determined at 03:38:53.6 UT on December 26th.   At 03:38:54.05 GPS time, one of the detectors of SEDA-FIB on board the International Space Station (ISS) detected a photon, arriving from the direction of Corona Australis near the Galactic Bulge (GB).   The energy of the photon was about $(35\pm7)$ MeV. A 3.0 σ level detection of the photon is discussed.

Introduction

The LIGO-Virgo team observed a signal of the Gravitational Wave (GW).   From the observed waveform, they determined the merging time of the two black holes at 03:38:53.6 UT on December 26th, 2015 (1).   In association with this event, by chance, one of the detectors of the Space Environment Data Acquisition System (SEDA) on board the International Space Station (ISS) recorded a photon.   In this paper we discuss whether the photon was coming from the diffuse gamma-ray source near the Galactic Bulge (GB) or in accordance with the merging process of two black holes.

Detector SEDA-FIB

The system called SEDA-FIB has been prepared for the measurement of the space radiation environment around the ISS.   The SEDA-FIB is composed of the scintillation fibers.   The detector was designed to measure the energy and the direction of neutral particles.   The scintillation fiber is quite wide at 6mm (wide)×3mm (thick)×98mm (long), so it may be adequately referred to as the scintillation tube.   One layer is made of 16 tubes forming a plane and those 16 planes are alternatively stacked, keeping right angle to each other to form the detector.

The detector is prepared for the observation of high-energy solar neutrons induced by solar flares as shown in Figure 1.   However the SEDA-FIB also has another function that is able to detect photons.   The energy of photons can be measured by converting an incident photon into an electron-positron pair and the direction of the incident photon can be determined by using the information of the tracks of electron and positron.

Electrons with energy of 3-30 MeV can be detected by the SEDA-FIB.   The energy of photons converted into an electron-positron pair is measured by the 256 channel ADC converter. The absolute value of the energy was calibrated by the proton beam at Riken (2).   The energy resolution is obtained as ~18% for protons with an energy of 35 MeV when we use the dynode out-put signal (2).   When we apply the range method, it is improved to 10%．In the present case the energy of the photon was measured by the dynode signal.   We also made a test run by using the photon beam with an incident energy of 100MeV and obtained the energy resolution for photon incidents.   The photon beam was provided by the INS electron accelerator with the electron energy of 350 MeV (3).   (By chance our experiment was the last experiment of the INS accelerator of the University of Tokyo.)

However in case that the converted electron or positron penetrates the internal volume of the scintillator and arrives at the anti-counter that surrounds the 6 surfaces of a cubic sensor, the event will not be recorded.   The anti-counter normally rejects those events.   Events are only recorded when the charged particles are involved in the 10cm×10cm×10cm internal scintillator block.   The anti-counter was prepared in order to record only neutral particles under enormous amount of background by charged particles.

The identification between neutrons and photons is easily made.   Neutrons are detected by converting them into protons.   Those low energy protons with the energy around 35-120 MeV lose a large amount of energy in each tube.   Especially at the end point of the track, they leave a dense Bragg peak.   The red spot in Figure 2 corresponds to the Bragg peak where the proton lost a large amount of energy by the ionization.   On the other hand, the ionization loss by either electrons or positrons is usually small and they are normally recorded as the minimum ionizing tracks.   The detection efficiency of photons can be calculated by using the radiation length of the scintillator, 44 cm.   In case we require a condition that photons must be converted into the electron and positron pair within 6mm from the surface of the scintillator block, the conversion probability is estimated as to be 0.011.   Technical details have been published in elsewhere (4 - 7).

Event Time

When the LIGO-Virgo collaboration found the gravitational wave, the ISS flew near the north of Easter Island. The position was 106.6°E, 19.7°S. The ISS was passing over the night region of the Earth. Fortunately the ISS was flying over the equatorial region where the background to the FIB is the lowest and the trigger rate was about 5 events per minute or one event per 12 seconds. The GPS time is provided to SEDA-FIB, so when we compare with the Universal Time, 17 seconds must be reduced from the GPS time. Therefore the GPS time 03:38:54.05 should be 03:38:37 UT. There is 17-second time difference between the SEDA and the LIGO-Virgo events, so within 17 seconds ~1.4 background particles may be triggered by the SEDA-FIB. If we consider only photon background, the chance detection with general photon background within 17 seconds is reduced as ~1/3 and as to be ~0.47 photons per 17 seconds (gate time). For the detection of solar neutrons, we can reduce the backgrounds to ~1/64 by using the directional information of the Sun but this time we cannot use this technique. The background is induced by the interaction between cosmic rays and the ISS frame. So up to this stage, the SEDA event seems to be consistent with the background induced by cosmic rays.

Arrival Direction

The arrival direction of the photon was from R.A. = 19h44m12s (or 296.0°) and declination $\delta$ = -38.4°. It is the direction of Corona Australis near the Galactic Center (R.A. 17h45.6m, $\delta$ =-28°56') and about 20 degrees northern part from the Galactic disk. A position error on the coordinate is estimated as to be 8 degrees. The energy of photons was (35±7) MeV. Therefore we try to estimate general background that may enter into an 8-degree opening cone. The ratio between the 8-degree open cone to a half sky region ($2\pi$ steradian) is given by $(8/57)^2/2 \approx 0.00975$. So we multiply this factor and reduce the above background (~0.47). Then we get a value of 0.47×0.00973=0.00457 (or 2.85σ). If the detected photon was really produced with the gravitational wave, the above probability may have an important scientific meaning.

In Figure 3a, we present the event display of the Y-Z plane. Figure 3b represents an interpretation for these tracks. From Figure 3, we notice that a track started to run from bottom to up in the sensor along the Y axis and it was separated into two charged tracks in the middle of the detector. This implies that a photon entered into the detector horizontally and converted into an electron and positron pair. After multiple Coulomb scattering, they are separated. A note is given on Figure 3 that the origin of the Y and Z axes is at the top-left corner and the Y-axis and Z-axis correspond to the horizon and vertical directions of ISS respectively. The

ISS flies over 400 km above the Earth, therefore we can see near the horizon without the effect of the atmosphere.

Flux of Diffuse Gamma-Rays

Since the arrival direction was not far away from the Galactic disk (~20 degrees north), we examine the possibility that the photon was originating from the diffuse gamma-rays. In fact, if there was no alert from the gravitational wave group, we simply regarded it as a component coming from a diffuse gamma-ray source of GB. The intensity of gamma-rays along the Milky Way has been measured by several satellites starting from the OSO-III to SAS-II, COS-B, CGRO, and FERMI. Those results except recent FERMI are summarized in the books of references 8) and 9). From the data of the bibliographies, we derive the intensity of gamma-rays around 30-40 MeV from the Galactic Center (GC) as to be $\sim 2 \times 10^{-3}$ photons/sec/cm$^2$. From the 20-degree northern region of the GC, the flux is expected to be one order less as to be $\sim 2 \times 10^{-4}$ photons/sec/cm$^2$.

Since the surface area of our sensor is 100 cm$^2$, ~ 0.2 diffuse gamma-rays are expected per detector per second from the GC and ~ 0.02 photons at 20-degree north from the galactic disk near the GC. Let us assume that the SEDA observed Galactic Bulge (GB) for 30 minutes. From now we will discuss the flux of diffuse gamma-rays based on the value of 0.02. Then within 30 minutes, we shall receive ~36 photons originating from the diffuse gamma-ray sources in the GB. However we must take account of the conversion probability of those photons inside the detector; it is estimated as ~ 0.011 (=1- exp (- 6mm/56cm)). Here the 56cm arises from the photon conversion length in the plastic scintillator (44cm×9/7) and 6 mm corresponds to the length where the incoming photons are converted into an electron-positron pair at the first layer. Then we actually observe ~0.40 photons per 30 minutes. The probability that the diffuse gamma-rays arrive within 17-second during the 30-minute observation time making a coincidence with arrival of the gravitational wave, may be small and it is estimated as to be 0.40×17sec/1800sec = 0.0038. This probability corresponds to ~2.9 σ. Those probabilities are listed up on Table I. If GW was produced at far beyond the Galactic Bulge, those statistical significance will have a scientific meaning.

Discussions Was this photon produced as a result of the merging process of the two black holes? Does the merging process associate with an intensive photon emission? In order to get the answer to these questions, we need another astronomical information near the merging region of the two black holes. After an observation, if no material will be left in the region except the gigantic black hole, no particle acceleration might happen. In fact, in this case we may not see any optical counterpart.

However in the opposite situation, if there was a magnetic field flowing into the black holes together with plasma, particle acceleration might happen.  The arrival time of the high energy gamma-rays is not necessarily required as the same time of the merging time of the two black holes.  By mutual interaction between the magnetic fields, if very strong electric field was instantaneously induced by $\boldsymbol{E}= -\partial \boldsymbol{B}/\partial t$, warm particles around the two black holes were accelerated into high energy, even beyond $10^{21}$ eV.  Kotera and Silk have discussed such a possibility that the merging process of giant black holes may be the origin of the highest energy cosmic-rays (10).

Those accelerated high-energy protons collided with the dense gas surrounding the merging region around ~100 pc and might produce strong gamma-ray beam.  Those gamma-rays were scattered away in the universe and collided with the materials either in the other galaxy or with the dense gas inside the GB of our Galaxy.  By such a *relay transmission* of gamma-rays, the 35 MeV photon might arrive near the Earth.

Finally we should mention about a report from the GRB monitor on board the ISS.  The CALET-GRB detector was looking at the zenith region at the same time with us and no positive signal was observed (11).  On the other hand, we observed not only the zenith but also the horizontal region, since we have been observing at the port #11 from September 30th, 2015.  We present both results on Figure 4.  (Figure 4 has been actually prepared by the CALET-GBM team).  So both observations are consistent.  The arrival direction of the photon observed by the SEDA-FIB detector is shown on the picture by the white circle.  It is near the GC.  The GC is marked by the cross-circle.  Here the area surrounded by the green line represents the most probable zone predicted by the LIBO-Virgo team where the two black holes merged.

Since we did not observe the sky by using the SEDA-FIB detector between May 20th and September 30th for the movement of the SEDA detector from Port #9 to Port #11, we missed to see the first event of GW150914.  However, the LIGO-Virgo team starts the observation again of the gravitational wave, so if the photon arrival was real in association with GW, once again in the next event we shall see it.

In summary, we detected a photon near the arrival time of the gravitational wave. There are three possibilities on this event: (1) a background photon induced by cosmic rays, (2) a photon coming from the Galactic Bulge, and (3) a photon in relation with the gravitational wave.  To confirm whether this photon originated from the cosmic ray induced or not, we must investigate more details about the frame of the ISS.  After the investigation, if the photon is surely coming from the space, from the direction of Corona Australis, there is a possibility that this photon is produced in association with the merging process of the two black holes.

Acknowledgements

  The authors thank to Goka T., Obara H. and Okudaira O. for various support for this project.

Table I.   The chance probability list

| Contents (combination) | Probability | σ |
|---|---|---|
| Chance coincidence with background | 0.40 | |
| Chance entrance of background-γ within narrow cone | 0.00457 | 2.85 |
| Coincidence with diffuse gamma-rays from GB | 0.0038 | 2.9 |

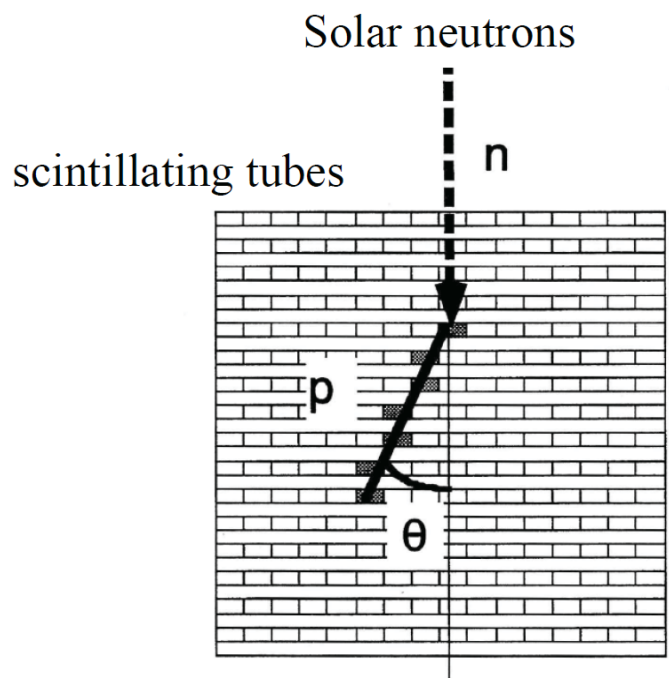

Figure 1. A schematic diagram of neutron detection by SEDA-FIB. Solar neutrons enter the sensor and converted into protons by a charge exchange interactions. The energy is measured by either tracking length or deposit energy. Actual example is shown in Figure 2.

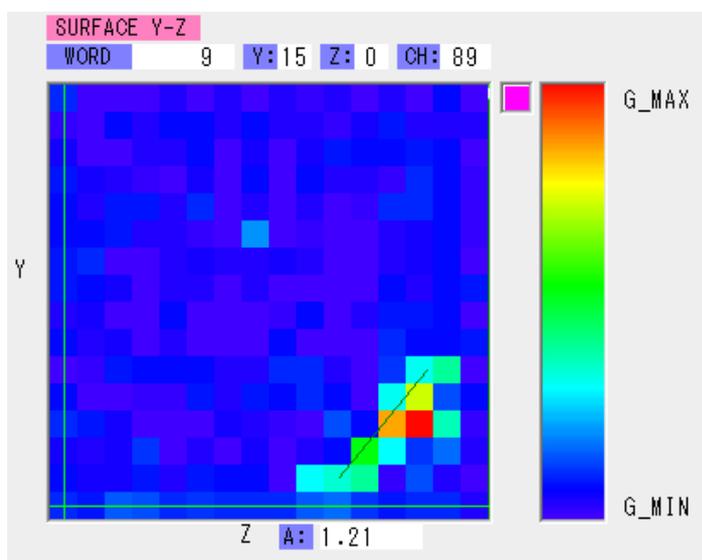

Figure 2. A typical example of a proton track. A neutron enters into the scintillator from the bottom side and converted into a proton inside the scintillation tube. At the final position of the running point, a Bragg peak is recorded (the red spot). The color corresponds to the ionization loss measured by the ADC counter. The G_min is set at the channel 3 of the ADC counts, while G_max is set at the channel 64. If you compare the track of Figure 3a, the difference is clear. The color scale is kept as the same for both pictures. In Figure 3a, no dense (red) ionization loss (point) is recognized. The minimum ionizing particles normally deposits its ionization peak around 17~25 channel, while the noise level is located around ~3 channel.

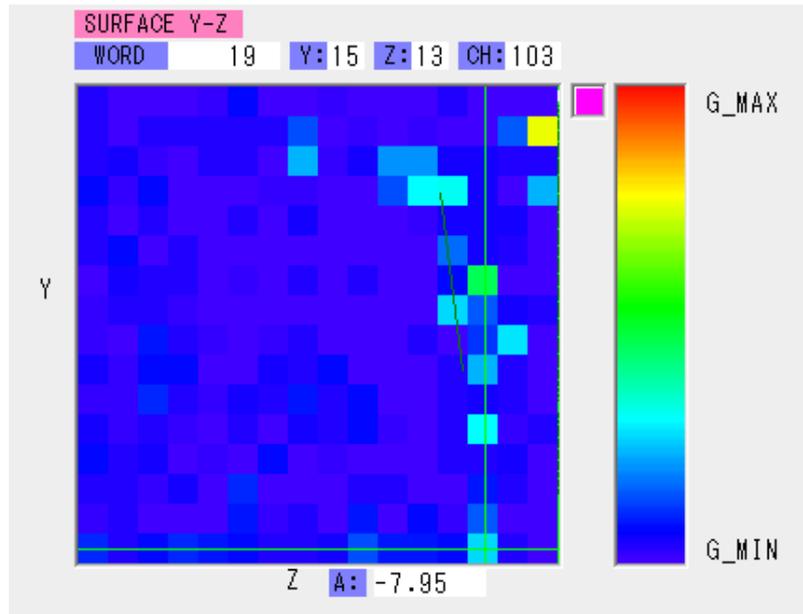

Figure 3a. The event display of the Y-Z plane. The origin of the Y-Z axes is at the top-left corner. From bottom to top, a thin tack is recognized. In the middle the track it is separated into plural tracks. They may correspond to an electron and a positron. The vertical line of Fig 3a is drawn to show how the photon entered horizontally. The vertical line corresponds to the horizon of the ISS. The G_min and G_max is set at channel 3 and 64 respectively.

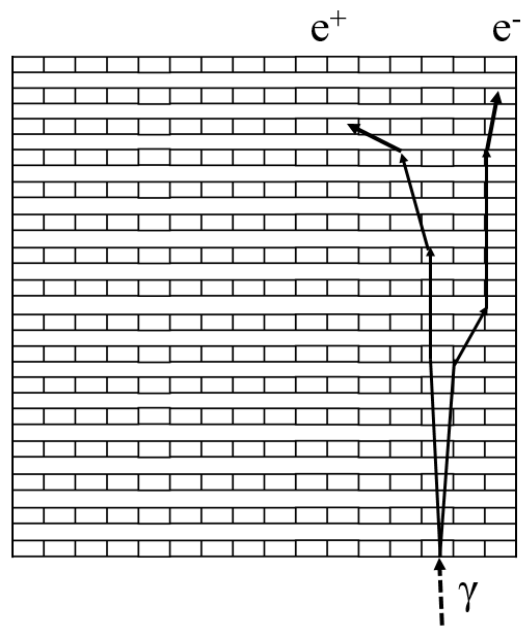

Figure 3b. An interpretation of the tracks. A photon incidents from the bottom side and converted into an e$^+$e$^-$ pair at the first layer of thickness 3mm. The electron and positron tracks can be identified by the scintillation tube from the middle part of the sensor.

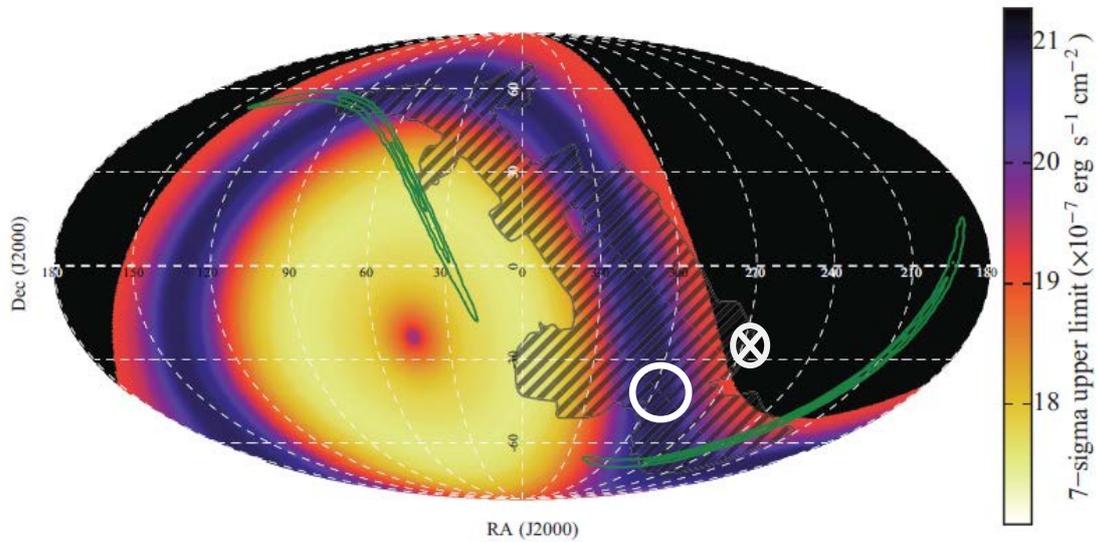

Figure 4. The sky map of the region where the CALET-GBM detector observed the soft gamma-rays at the time of GW151226 (negative result). The ISS flew over the South hemisphere over 106.6°E and 19.7°S, and the most sensitive region for CALET-GMB is shown by the bright orange color. The detection region of the photon by the SEDA-FIB detector is shown on the map at the down side of the South hemisphere by the white circle (RA=296°, δ=-38.4°). The green belt represents the region where the merging process of BH predicted by the LIGO-Virgo. The galactic Center is shown by the white cross circle.